\documentclass[superscriptaddress,onecolumn,amssymb,amsmath,nobibnotes,
aps,prd,nofootinbib]{revtex4}
\pdfoutput=1
\usepackage{graphicx,subfigure,bm,color,hyperref}
\usepackage{amsfonts}
\usepackage{verbatim}
\usepackage{color}
\begin{document}

\title{Thermodynamics and Phase Transition in Shapere-Wilczek {\it fgh} model:\\ Cosmological  Time Crystal in Quadratic Gravity} 
	
\author{Praloy Das}
\email{praloydasdurgapur@gmail.com}
\affiliation{Physics and Applied Mathematics Unit, Indian Statistical Institute, 203 Barrackpore Trunk Road, Kolkata-700108, India}
	
\author{Supriya Pan}
\email{supriya.maths@presiuniv.ac.in}
\affiliation{Department of Mathematics, Presidency University, 86/1 College Street, Kolkata-700073, India}
	
\author{Subir Ghosh}
\email{subirghosh20@gmail.com}
\affiliation{Physics and Applied Mathematics Unit, Indian Statistical
Institute, 203 Barrackpore Trunk Road, Kolkata-700108, India}

\begin{abstract} 
	The Shapere-Wilczek model \cite{wil}, or so called   $ fgh$ model, enjoys the remarkable features of a Time Crystal (TC) that has a non-trivial time dependence in its lowest energy state (or  the classical ground state). We construct a particular form of $ fgh$ model (with specified $f,g,h$ functions)  that is derived from a
	Mini-superspace version of a  quadratic $f(R,R_{\mu\nu})$ gravity theory. Main part of the investigation deals with thermodynamic properties  of such systems from classical statistical mechanics perspective. Our analysis reveals the possibility of a {\it phase transition}.  Because of the higher (time) derivative nature of the model computation of the partial function is non-trivial and requires  newly discovered techniques. We speculate about possible connection between our model and the Multiverse scenario.

\end{abstract}	

\maketitle
	
{\bf{Introduction:}} 
The classical and quantum versions of the Time Crystal (TC) have generated an enormous amount of interest just within a few years of its theoretical possibility, conceived by Shapere and Wilczek \cite{wil} and by  Wilczek \cite{wil1} respectively. A similar idea (leading to a spatially varying ground state condensate) in a different framework was also proposed by one of the present authors in \cite{sg}.  (For a recent review and references, see \cite{rev}.) 
 After the critical assessment of \cite{bruno} of the original quantum version given in \cite{wil1}, there has been major theoretical developments \cite{th} of the quantum TC with remarkable experimental verifications \cite{ex}. On the other hand the classical version of TC is free from controversies \cite{wil2}. However, the Classical TC (CTC) is comparatively less studied mainly due to lack of realistic models (however see \cite{wil3} for an explicit classical model).  In the cosmological context, a relativistic scalar field with a non-canonical kinetic term, in an expanding Friedmann-Robertson-Walker (FRW) universe, induces a TC behavior \cite{Bains}.  More recently, we have shown \cite{pra} that TC behavior in a noncommutative extended FRW model \cite{stern} where the scale factor, (being   the only dynamical variable), exhibits periodic behavior indicating the possibility of a bouncing universe \cite{cycl}. It is interesting specially because the bouncing behavior is achieved in a purely geometric setup, without any matter field. TC in (cubic) $f(R)$ form of gravity has been proposed in \cite{feng}. The present work  deals with a specific form of quadratic $f(R,R^{\mu\nu}R_{\mu\nu})$ gravity but our main focus is on its thermodynamic aspects from a statistical mechanical perspective.

 In the theme of cosmology in TC scenario a recent important work is by Vacaru \cite{vac}. The author has considered in detail TC behavior  in generic off-diagonal,
 locally anisotropic and inhomogeneous metrics in a modified gravity framework, by exploiting earlier works in this context \cite{vac1}. The motivation of studying extended gravity models stems from the fact that observational results of Planck$2015$ \cite{planck} suggest that  conventional and minimal models  of inflationary cosmology are not sufficient for a  unified description of
 inflation with dark energy era. In particular the author in \cite{vac} has analysed TC in Starobinsky quadratic gravity model $f(R,R^2)$ \cite{star} that agrees with the Planck$2015$  data \cite{planck}. It should be noted that, even though  the premises of   the work in \cite{vac} and the present work is somewhat similar, the focus of the two studies are entirely different. In \cite{vac} the author has considered purely cosmological effects in a generalized spacetime dependent metric  and only at the end restricts to Starobinsky model \cite{star}. On the other hand we have taken up (a slightly extended version $f(R,R^{\mu\nu}R_{\mu\nu})$ of) Starobinsky model in FRW framework and have concentrated on its thermodynamic features from a statistical mechanical perspective.

It is now established that (classical and quantum) TC is a novel and distinct {\it {phase}}. This immediately raises  questions about its thermodynamic behavior.  It is probably debatable whether conventional equilibrium statistical mechanics can be applied to TC that are curiously in the borderline between  equilibrium and non-equilibrium system. Since we have an unambiguous mechanical (albeit non-canonical) particle model for CTC, we forge ahead to construct its partition function and come up with the remarkable result of a possible {\it {phase transition}} in the system. Computation of the partition function exploits newly established techniques: higher derivative nature of the parent model leaves its signature in a slightly modified version of the model (to avoid ghosts), following the scheme proposed in \cite{tolly} and integration  for computing the partition function requires a non-trivially modified measure \cite{zhao} (due to the presence of quartic velocity term).\\
{\bf {Classical (generic) Time Crystal:}} The spontaneous symmetry breaking paradigm is pivotal in the general  phenomenon of crystallization. In its conventional manifestation ground state (or minimum energy state) consists of  the atoms arranging themselves   into a definite periodic lattice throughout the configuration space, thereby breaking spontaneously the spatial translation symmetry. In \cite{sg} higher spatial derivative terms  induced a spontaneous symmetry breaking in momentum space leading to the lifting of translational invariance. In \cite{wil,wil1} examples of models were provided whose ground state is endowed with a periodic motion leading to a Time Crystal that violates time translation symmetry. Note that  Hamiltonian dynamics forbids the existence of  CTC. For a Hamiltonian $H(p,\phi)$ with coordinate $\phi$ and conjugate momentum $p$,  $H(p,\phi)$ minimizes at $\frac{\partial H}{\partial p}=\frac{\partial H}{\partial \phi}=0$. But   Hamilton's equations of motion states  $\dot\phi = \frac{\partial H}{\partial p}$. Put together we get for the minimum energy state $\dot\phi = \frac{\partial H}{\partial p}=0$ indicating that $\phi $ should be a constant. Thus classical ground state should be static contrary to a CTC ground state. This negative conclusion can be bypassed if the structure of $H$ is such that the
	 canonical momentum $p$ leads to a multivalued Hamiltonian as a function of $\dot \phi$ with cusps at $\frac{\partial p}{\partial \dot \phi}=0$ where the Hamiltonian equations of motion are not valid. In the CTC models of  \cite{wil,wil1, pra}   the system ground state has to  adjust itself to the contrasting demands of a time invariant (constant position) and simultaneously time varying (constant velocity) state.\\
{\bf {Classical (Cosmological) Time Crystal:}} In the present paper we study a generalized form of gravity, popularly known as $f(R_{\mu\nu})$ gravity with up to quadratic invariants made out of the Ricci tensor and scalar,  
  \begin{eqnarray}
  \mathcal{A} = \frac{c^4}{16 \pi G} \int \sqrt{-g} (R+ C R^2+D R^{\mu\nu} R_{\mu\nu}- 2 \Lambda) \; d^4x 
  \label{aa}
  \end{eqnarray}
  where $C,D$ are numerical constants.  In the cosmological context, adhering to the  minisuperspace formalism, the action (\ref{aa}) reduces to,  
  \begin{eqnarray}
  \mathcal{A} = \frac{3 c^4}{2G} \int ~dt \Bigg [ \left( -a \dot{a}^2  + k a - \frac{\Lambda}{3} a^3 \right) + p \left(\frac{\dot{a}^4}{a}+\frac{k^2}{a} \right)+q (\dot{a}^2 \ddot{a}) +p (a \ddot{a}^2)+2pk \frac{\dot{a}^2}{a}  \Bigg ] \label{action}
  \end{eqnarray}
  where  $p=6C+2D$ and $q=12C+2D$. The $q$-term drops out being a total derivative. This Lagrangian is a variant of  the $fgh$ form of \cite{wil}, $L=f(a)\dot a^4 +  g(a)\dot a^2 + h(a)$, having in addition $\ddot a$ terms.

\begin{figure}
	\begin{center}
		\includegraphics[width=0.4\textwidth]{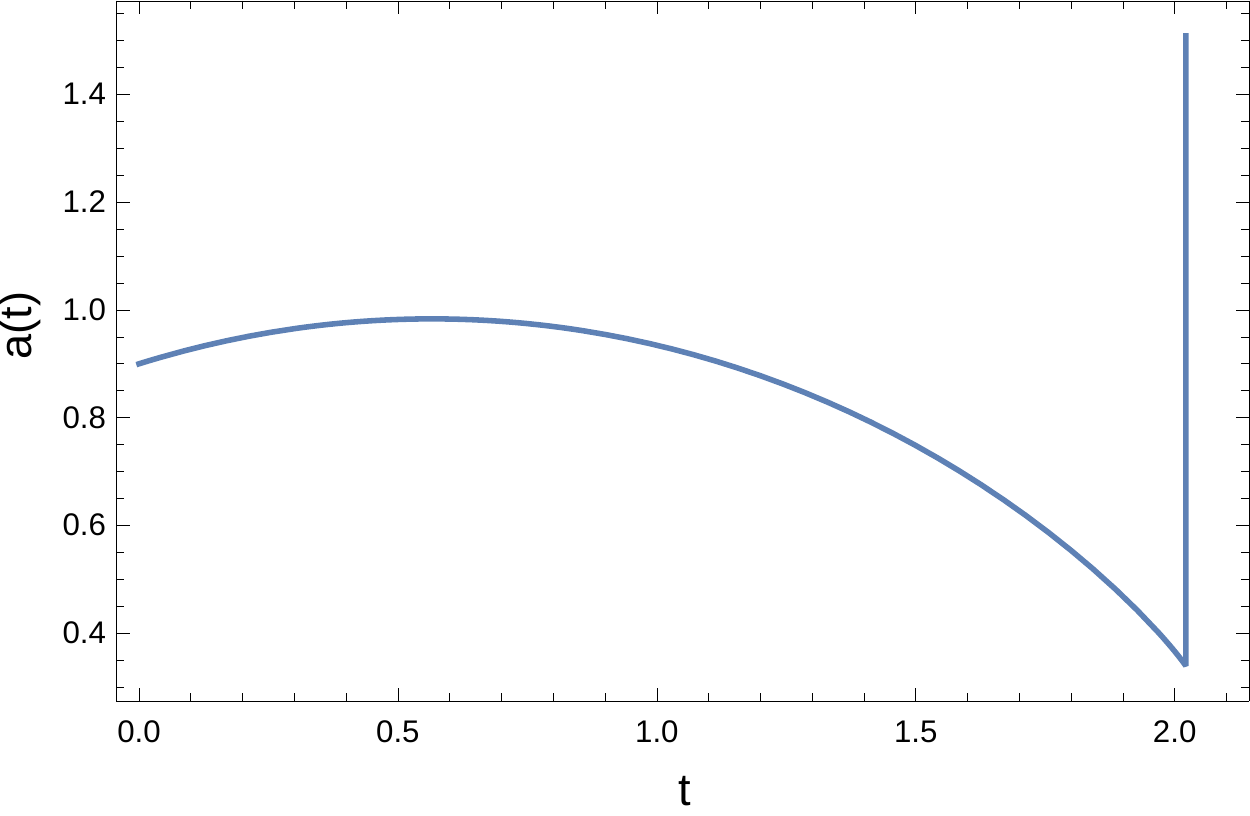}
		\caption{The plot for the scale factor $a (t)$, vs. $t$ showing the bounce and singularity during the evolution of universe for the master equation (\ref{motion}). }
		\label{fig1}
	\end{center}
\end{figure}

  This action has a complication since it consists of higher (time) derivative terms and will be plagued by ghost problems. Recall that in \cite{pra} we had dropped higher derivative terms, (present in the noncommutative extended FRW model \cite{stern}), in an approximation scheme. Incidentally, $f(R)$ models constructed out of Ricci scalar $R$ only are free from the ghost menace. For the mechanical models, (as the present one), the ghost problem manifests itself via a non-positive definite Hamiltonian. We follow the scheme of \cite{tolly} to develop a reduced model with a positive definite Hamiltonian (see Appendix  A for details). This method allows us to get rid of the higher time derivative term  thereby yielding 
\begin{eqnarray}\label{lag}
L=\left( -a \dot{a}^2  + k a - \frac{\Lambda}{3} a^3 \right) + p \left(\frac{\dot{a}^4}{a}+\frac{k^2}{a} \right)-\sigma \frac{\dot{a}^4}{4a}+2pk \frac{\dot{a}^2}{a}.  
\end{eqnarray}
From the Lagrangian (\ref{lag}), the equation of motion can be written as,
\begin{eqnarray}
\ddot{a}\left[-2a +12p\frac{\dot{a}^2}{a}+4p\frac{k}{a}-3\sigma\frac{\dot{a}^2}{a}\right]-\frac{3\dot{a}^4}{a^2}\left(p-\frac{\sigma}{4} \right)-2pk\frac{\dot{a}^2}{a^2}+p\frac{k^2}{a^2}-k+\Lambda a^2=0.
\label{motion}
\end{eqnarray}
We provide a plot in Fig. \ref{fig1} depicting  evolution of the scale factor $a(t)$ with the cosmic time $t$ by solving the master equation eqn. (\ref{motion}) numerically, assuming  spatial flatness of the universe (i.e., for $k=0$). The figure shows that at a certain time $t$, the scale factor shows a single hump before it hits the singularity at around $t \sim 2$. This phenomenon  can be interpreted as a cosmological bounce. 

Defining the canonical momentum as
\begin{equation}
p=\frac{\partial L}{\partial \dot a}=-2a\dot{a}+4p \left(\frac{\dot{a}^3}{a}+k\frac{\dot{a}}{a} \right)-\sigma\frac{\dot{a}^3}{a},
\label{p}
\end{equation}
the Hamiltonian reads,
\begin{eqnarray}
H &=&p\dot a - L  = 3 \left(p-\frac{\sigma}{4}\right)\frac{\dot{a}^4}{a}+2pk \frac{\dot{a}^2}{a}-a \dot{a}^2-p \frac{k^2}{a}-ka+\frac{\Lambda}{3}a^3 \nonumber \\
 &=& \frac{ 3(p-\frac{\sigma}{4})}{a}\left[\dot{a}^2-\frac{(a^2-2pk)}{6(p-\frac{\sigma}{4})}\right]^2 +V_{eff}
\label{h}
\end{eqnarray}
where the effective potential $V_{eff}=\frac{\Lambda}{3}a^3-p \frac{k^2}{a}-ka-\frac{(a^2-2pk)^2}{12a(p-\frac{\sigma}{4})}$. Similar ideas using spatial derivative have been employed in \cite{sg,dan}\\
{\bf{Minimizing $H$ to show Time Crystal behavior:}} In our model $a$ being  the scale factor is always positive. Hence renaming $(p-\sigma /4)\sim p> 0$, notice that $H$ can be minimized by separately minimizing the kinetic term, 
$\dot{a}_0^2=\frac{(a_0^2-2pk)}{6p}$ with $a_0$ determined from minimizing $V_{eff}$, i.e., $\partial V_{eff}/\partial a =0$. Thus, as long as $a^2_0-2pk > 0$, the system can behave as a CTC since the minimum energy state or ground state requires a non-zero velocity ($\dot a_0\neq 0$) as well as a non-zero coordinate ($ a_0\neq 0$).

{\bf{Thermodynamics of Classical Time Crystal:}} Thermodynamic properties of a generic CTC from a statistical mechanics perspective is the new element in our work where we utilize   the CTC Hamiltonian.  Thermodynamic features of this new phase of matter $-$ CTC $-$ have not been studied earlier and there are novelties both in computational procedure and in results. Very interestingly, we find indications of  phase transition as $\dot a$ increases from a small value to $\dot{a}_0=\pm {\sqrt{\frac{(a_0^2-2pk)}{6p}}}$ when the CTC phase sets in.

Although we have managed to replace higher time derivatives still the Lagrangian contains quartic term in $\dot a$ that is in fact necessary for the TC behavior. However this makes the conjugate momentum $p$ multiple valued resulting in different Hamiltonians. Conventionally one computes the partition function summing or integrating over phase space degrees of freedom, i.e., coordinate and momentum but for a generic CTC model this is not convenient.  However coordinate ($a$) and velocity ($\dot a$) prove to be the proper choice of variables, as advocated in \cite{henn}. The price to pay for this is the following: using $a,~\dot a$ instead of $a,~p$ is effectively a change of variables that brings in a Jacobian $J(a,\dot a)$ in the measure (see Appendix B for details of computing $J$ in our model). This has been discussed with specific examples similar to our model in \cite{zhao}. 
Hence, the partition function is (with the inverse temperature  $\beta =1/{k_BT}\equiv 1/T$  where the Boltzmann constant $k_B=1$),
\begin{eqnarray}
Z=\int _{\dot a=-\infty}^{\dot a=\infty}d\dot a~\int _{ a=0}^{ a=A}d a~J(a,\dot a)e^{-\beta H(a,\dot a)}.
\label{z1}
\end{eqnarray}
 Although the $\dot a$-integral is analytically doable  it is not possible to perform the $a$-integral analytically. Hence we take recourse to numerical integration. Our aim is to consider a canonical ensemble and compute thermodynamic potential such as the Helmholtz free energy $F$ and subsequently other thermodynamic observables such as average energy, entropy, and higher derivatives such as specific heat and compressibility. Our scheme is the following: we numerically calculate (\ref{z1}) for a set of different values of $A,~T $ (remember that ``volume'' is linear dimension $a$ in our case) and thus generate a set of values for $F$.  Note that factor $J$ in measure can become negative thus rendering $Z$ unphysical. This forces us to restrict the upper limit of $a$ {\footnote{ Note that restrictions on the dimension of the system appears  for a real gas as well. In van der Waal equation of state the volume   can not be smaller that the total volume of gas particles.}}.

In order to study rest of the thermodynamic quantities, analytic forms of $F(T)$ for fixed $a$ and $F(a)$ for fixed $T$ are obtained by curve fitting. Comparison with ideal one dimensional gas profile
 $$F_{\rm Ideal} = -T \left({\rm Constant} + \ln (A) +\frac{1}{2} \ln (T) \right),$$
it is remarkably clear that the CTC has two distinct type of behaviors: in one sector it resembles the ideal gas whereas in the other sector it behaves differently. In all the figures containing red and blue profiles, red line shows ideal gas behavior and blue line depicts the CTC behavior. This is revealed in Fig.(2) in a sort of phase diagram where the set of blue points for $P,A,T$ (which are computed later in the paper) represent CTC and the set of red points $P,A,T$ represent ideal gas, following the ideal gas equation of state $Pa=T$. Note the sharp step in the blue CTC surface for a fixed $A$ after which the blue (CTC) and red (ideal gas) surfaces follow similar behavior. As explained before the blue CTC surface does not continue beyond a specific value of $a$ ($a\sim 3.8$ for our model).
 \begin{figure}
	\begin{center}
\includegraphics[height=0.3\textwidth]{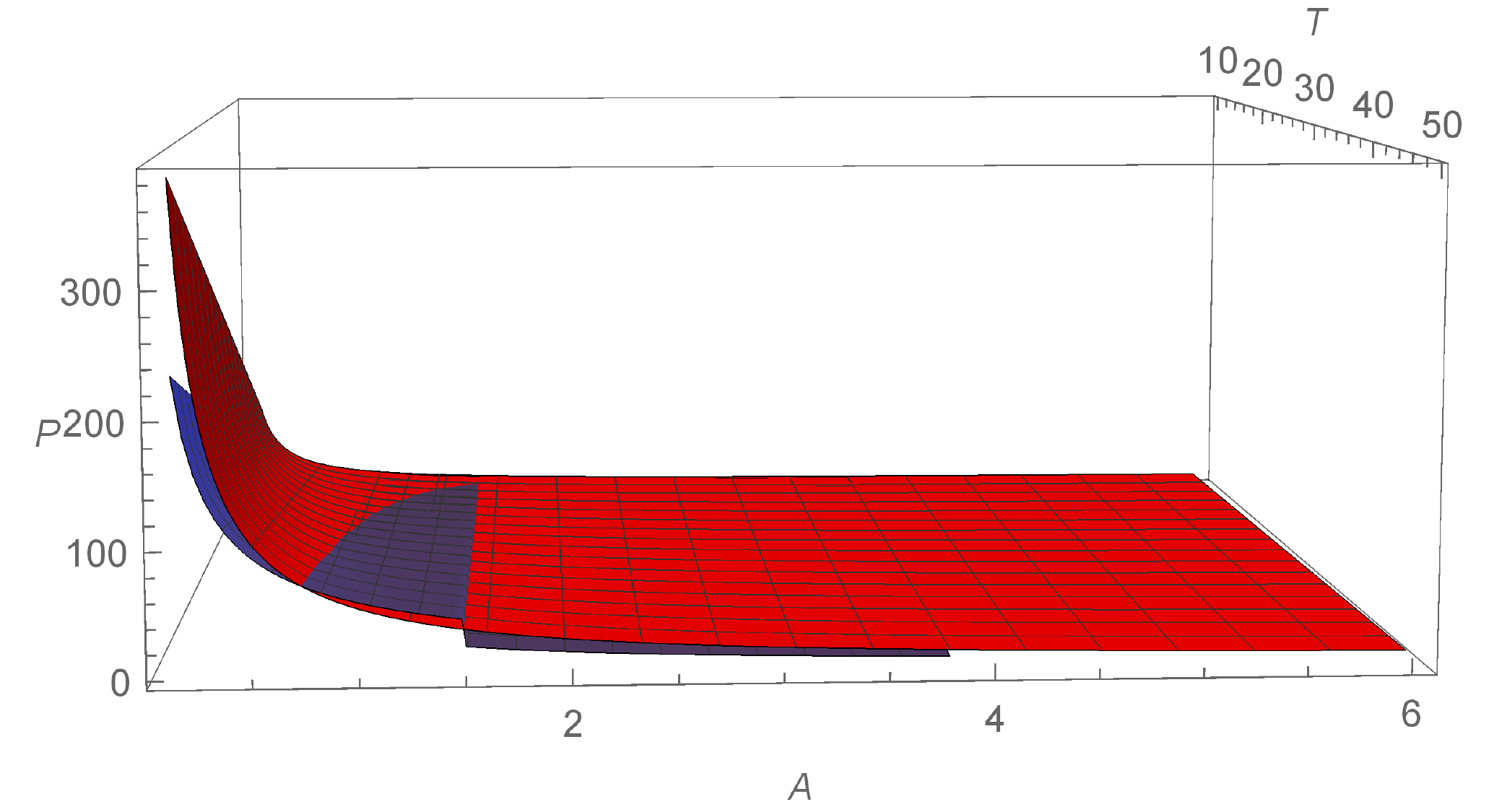}
		\caption{Phase diagram of CTC (blue surface) with bounded $A$ and ideal gas (red surface) with unbounded $A$. The two surfaces are similar for $A>3.8$ where the CTC (blue) surface exhibits a discrete jump. }
		\label{fig13}
	\end{center}
\end{figure}

\begin{figure}
  \begin{center}
 \includegraphics[width=0.4\textwidth]{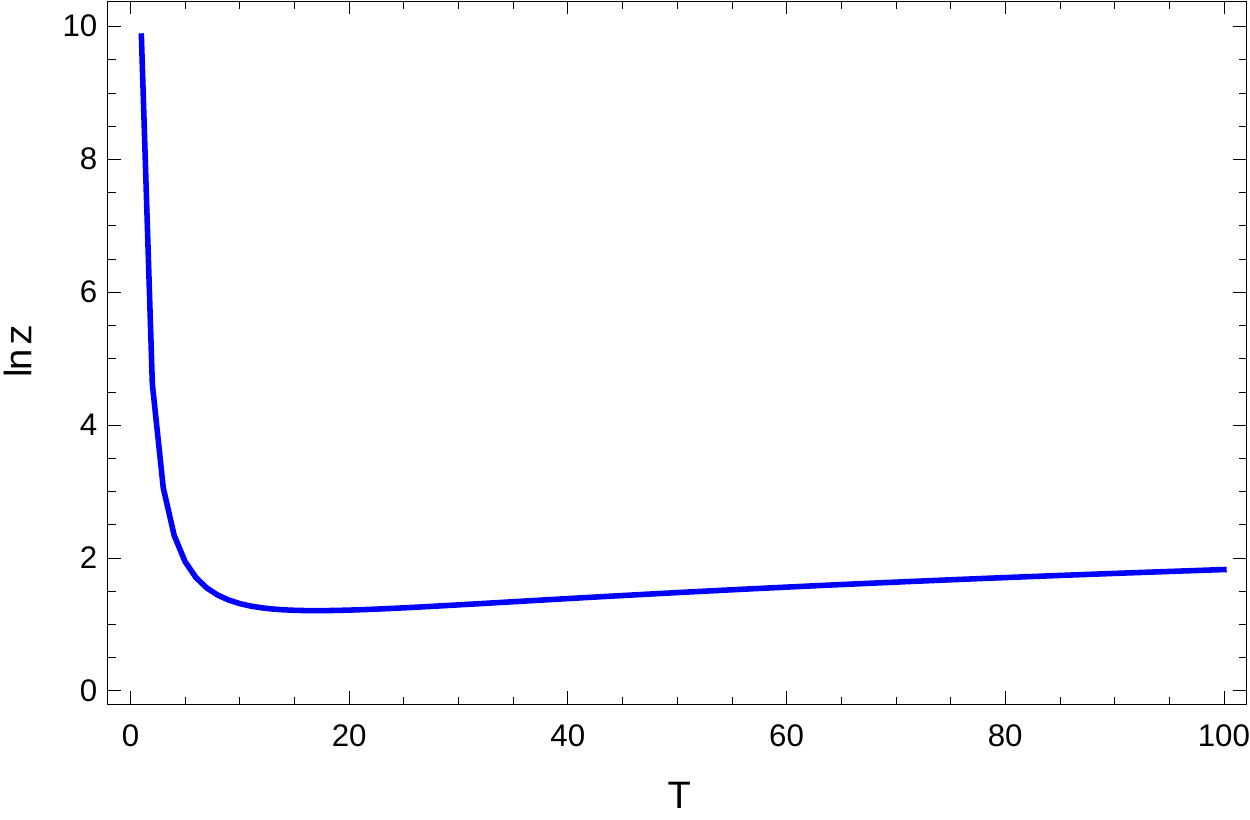}
 \includegraphics[width=0.4\textwidth]{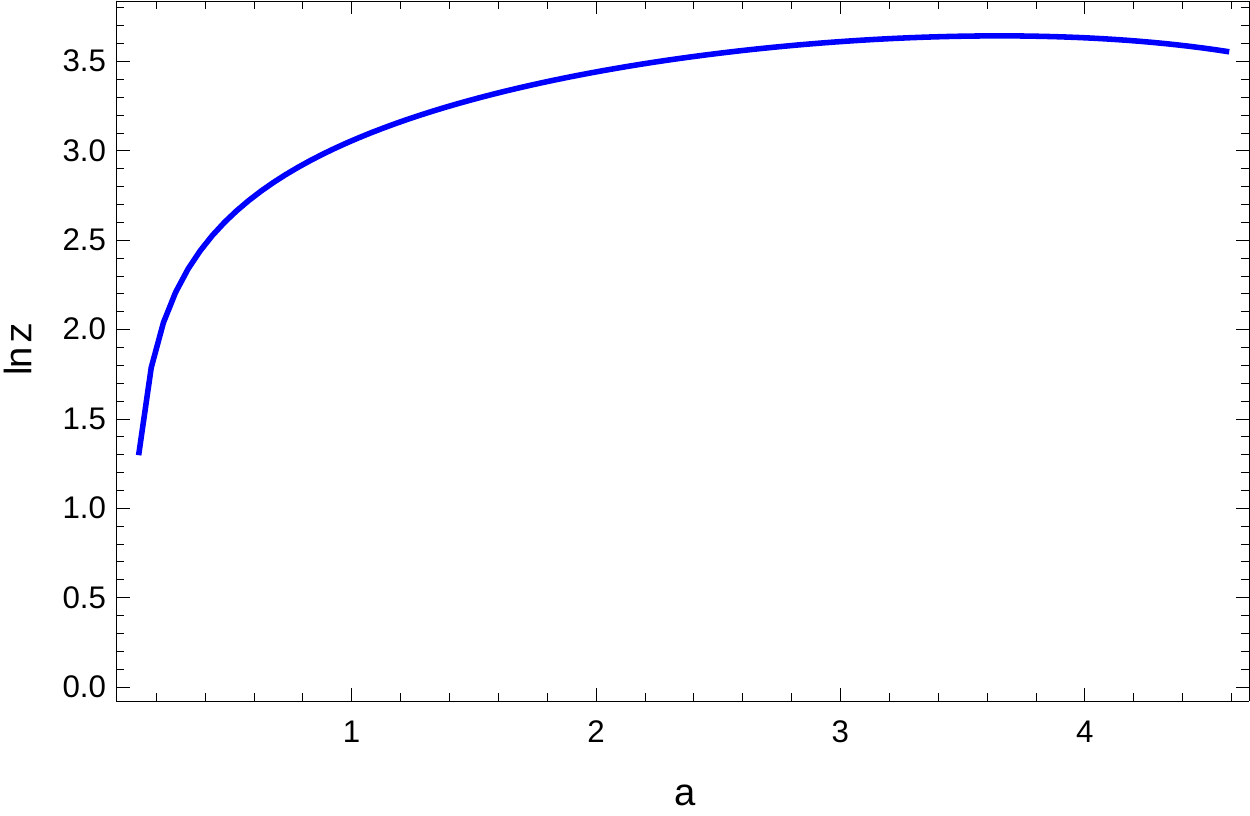}
  \caption{The left panel shows the qualitative evolution of $\ln~Z$ with respect to $T$ for the fixed scale factor $a$ while in the right panel the evolution of $\ln~Z$ vs. $a$ for the fixed $T$  has been displayed.}
  \label{fig3-4}
\end{center}
     \end{figure}

\begin{figure}
 \begin{center}
  \includegraphics[width=0.4\textwidth]{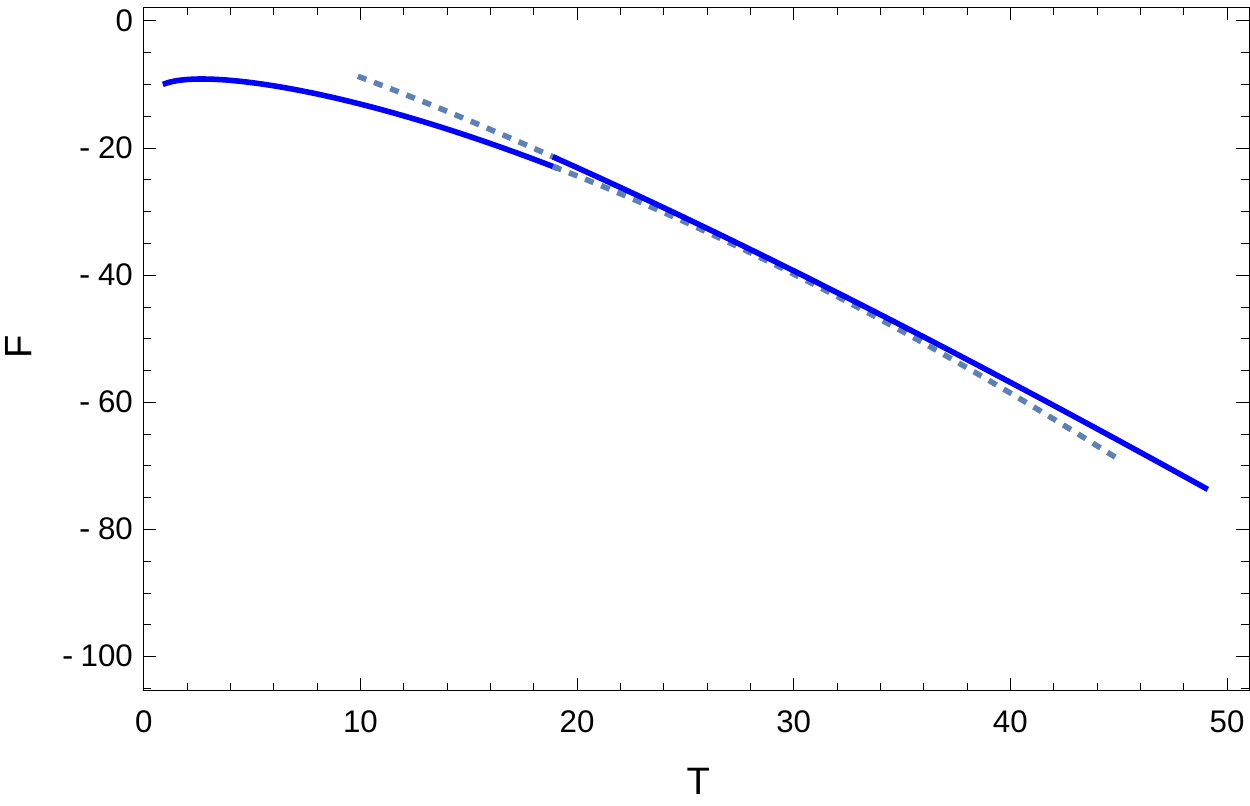}
  \includegraphics[width=0.4\textwidth]{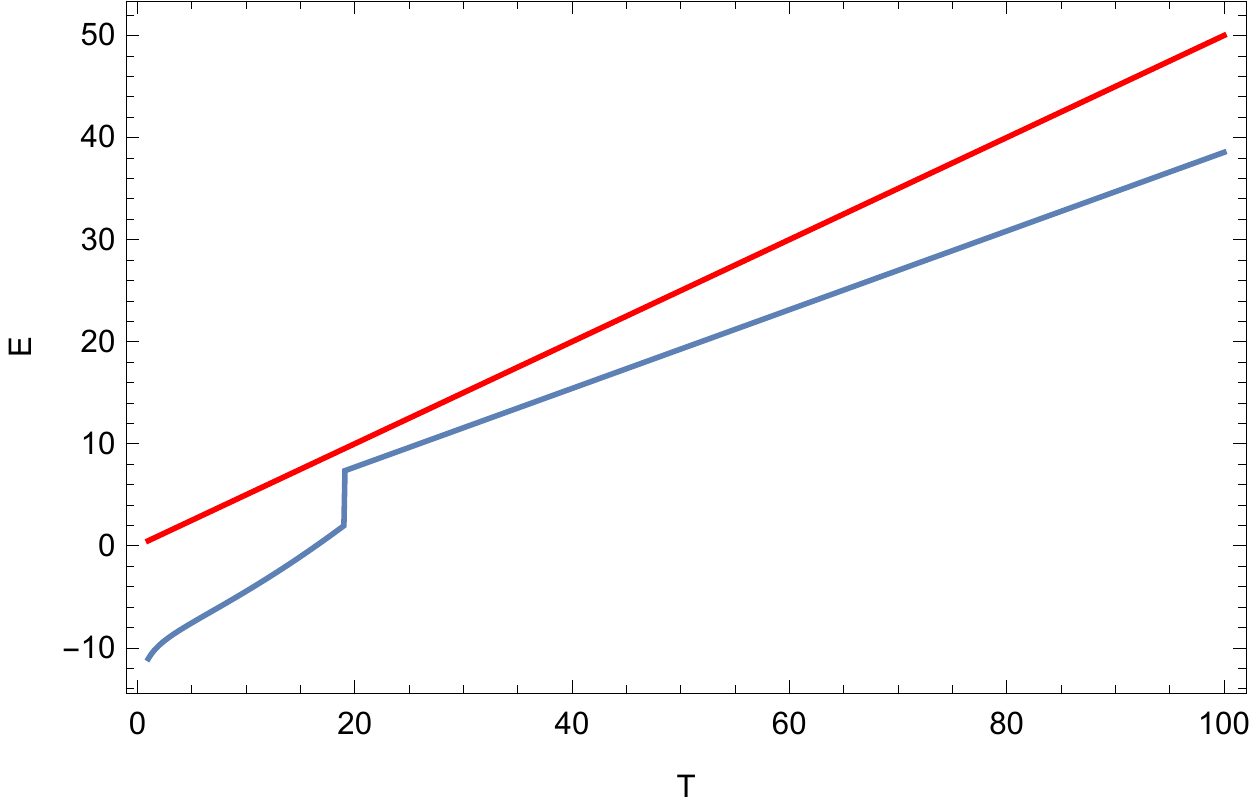}
  \includegraphics[width=0.4\textwidth]{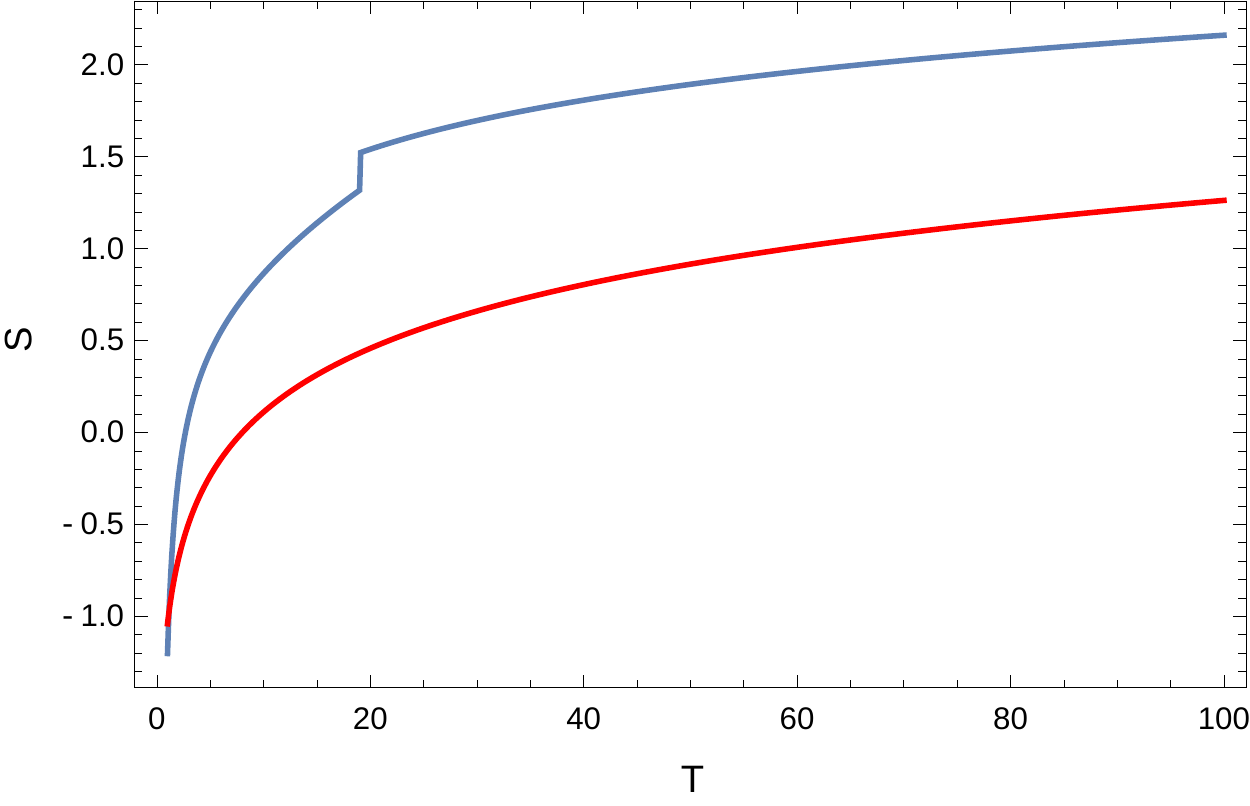}
  \includegraphics[width=0.4\textwidth]{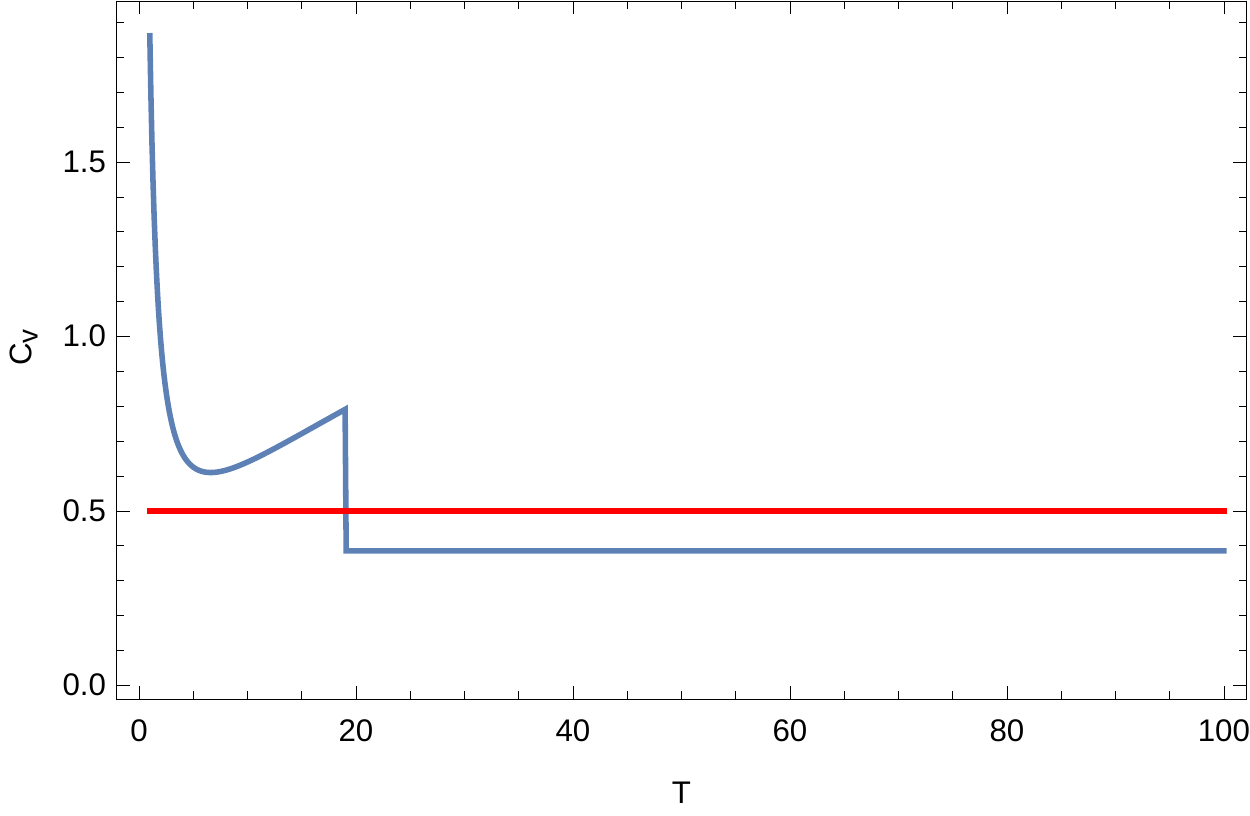}
 \caption{The figure presents evolutions of different thermodynamical quantities for the present scheme. {\it Upper left:} Plot of $F$ vs. $T$ with metastable states shown by dotted lines. {\it Upper right:} Energy $E$ vs. temperature $T$ plot showing discontinuity in first derivative of $F$. {\it Lower left:} Entropy $S$ vs. temperature $T$ plot showing discontinuity in first derivative of $F$. {\it Lower right:} Plot of specific heat $C_a$ vs. $T$ showing a break from continuity. In the plots the red and blue curves respectively represent the ideal gas and CTC profiles. }
\label{fig:5-8}
\end{center}
\end{figure}

 Our method is illustrated in Fig. \ref{fig3-4} where in the left panel of  Fig. \ref{fig3-4} we show the evolution of $lnZ$ as function of $T$ for fixed $a$ and in the right panel of Fig. \ref{fig3-4} we show the evolution of $lnZ$ as a function of $a$ for fixed $T$.  The qualitative difference in $lnZ$ above and below the transitional region requires two different analytic forms for $lnZ$ that we fix by curve fitting in {\it Mathematica }.
Notice that the CTC curve becomes similar to  the ideal gas curve in Fig. \ref{fig3-4} after a critical value of $T$ (or $a$) whereas the  CTC curve profile is qualitatively different below the critical region. To faithfully represent the numerical results it seems natural to fit the CTC profiles similar to ideal gas for regions above the critical value and as arbitrary power law for region below the critical value. In a set of plots given in Fig. \ref{fig:5-8}, 
we consider a prototype profile with $a$ fixed and plot Free Energy $F(T)=-T \ln Z$ (upper left panel of Fig. \ref{fig:5-8}), the entropy $S(T)=-(\partial F)/(\partial T)$ (upper right panel of Fig. \ref{fig:5-8}), energy $E(T)$ shown in the lower left panel of Fig. \ref{fig:5-8}, and finally the specific heat  $C_a(T) =(\partial E)/(\partial T)=T(\partial F)/(\partial T)-F$ (analogous to $C_V$) in the lower right panel of Fig. \ref{fig:5-8}. 
                        
It is interesting to note that with our fit for $\ln Z$ as function of $T$, $F=-T \ln Z$ is {\it {continuous}} across the critical value of $T$ with existence of metastable states along extension along both sides of the critical value of $T$ in the upper left panel of Fig. \ref{fig:5-8}. These are shown as dotted red and blue points  extending the ideal gas and CTC respectively. However, {\it  discontinuities do show up in the first and second derivatives of} $F$ {\it at a critical} $T$. The jump in $E(T)$ (upper right panel of Fig. \ref{fig:5-8}) and in  $S(T)$ (lower left panel of Fig. \ref{fig:5-8})  indicate possibility  of a {\it {structural}} change with a latent heat given by $L=T\Delta S$. This is further corroborated in the finite gap in $C_a$ profile against $T$ displayed in the lower right panel of Fig. \ref {fig:5-8}. 
                 
\begin{figure}
\begin{center}
\includegraphics[width=0.4\textwidth]{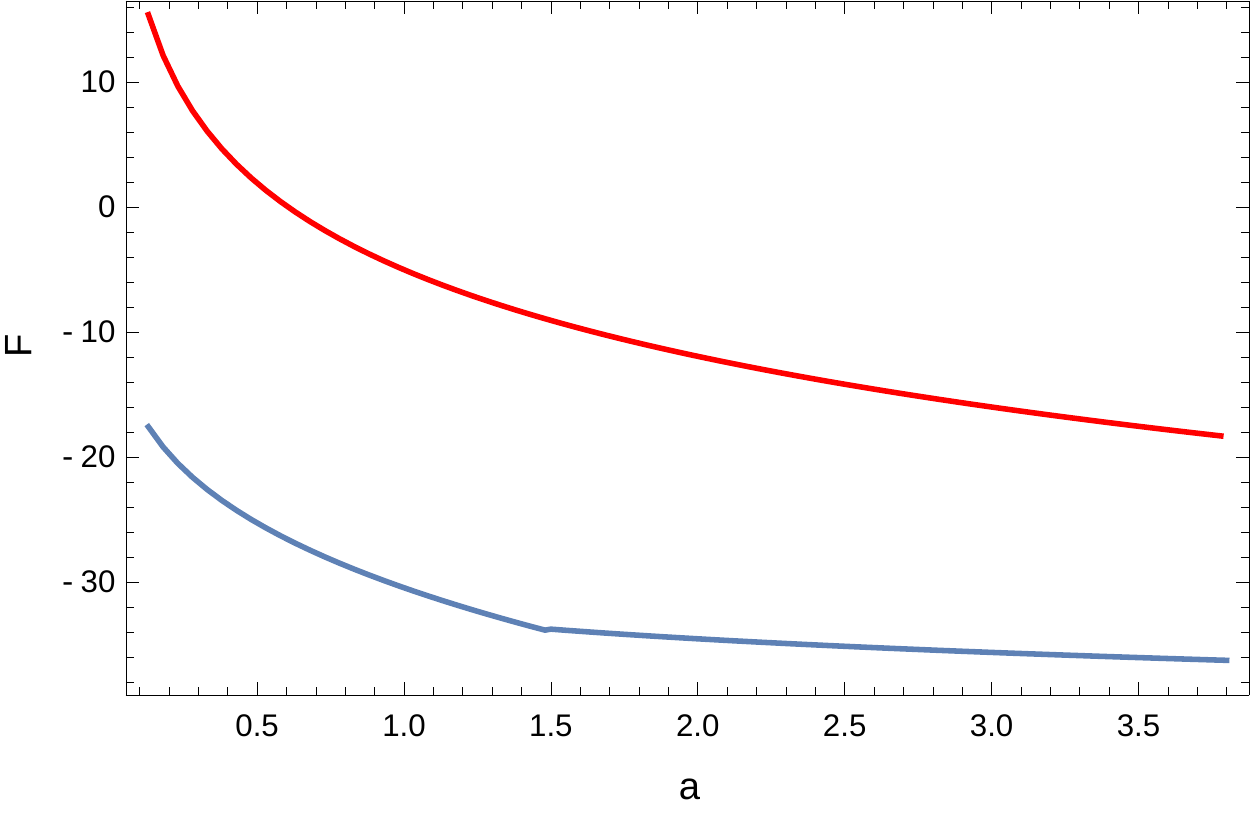}
\includegraphics[width=0.4\textwidth]{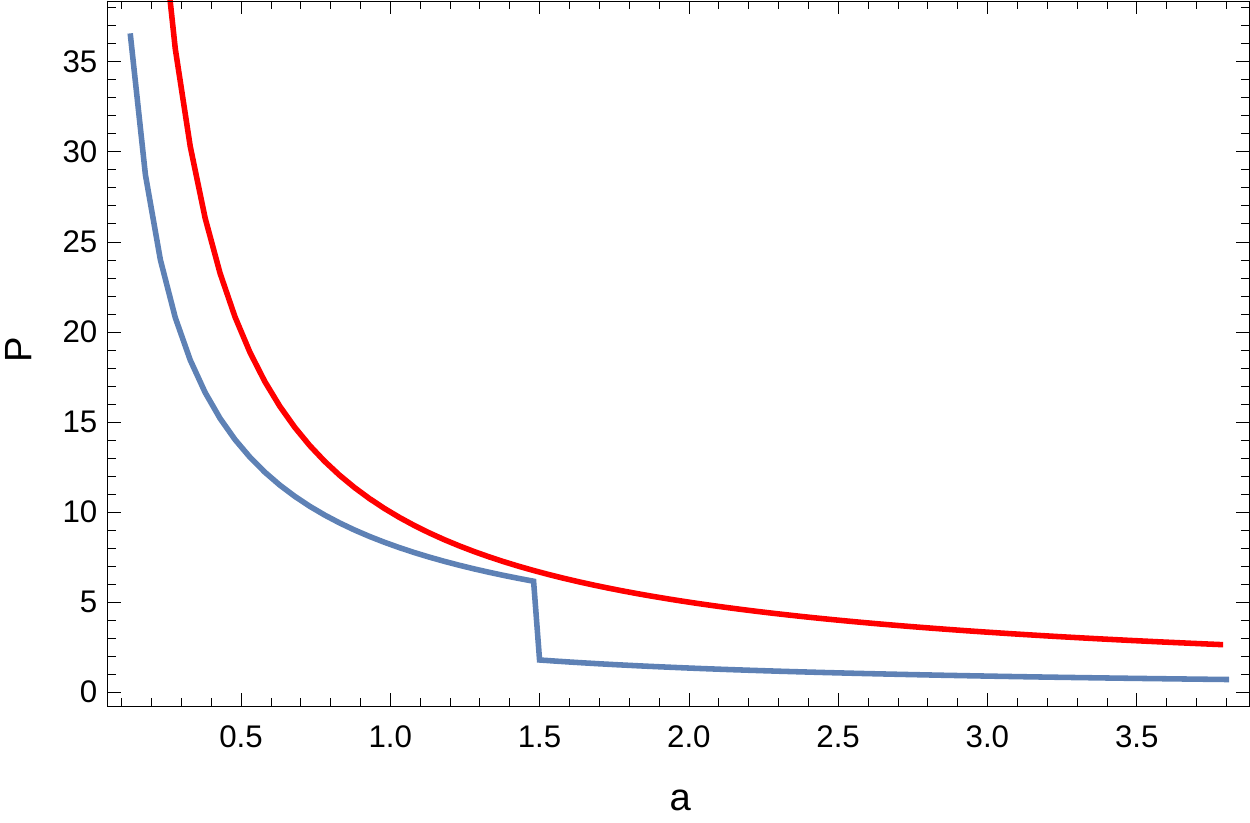}
\includegraphics[width=0.4\textwidth]{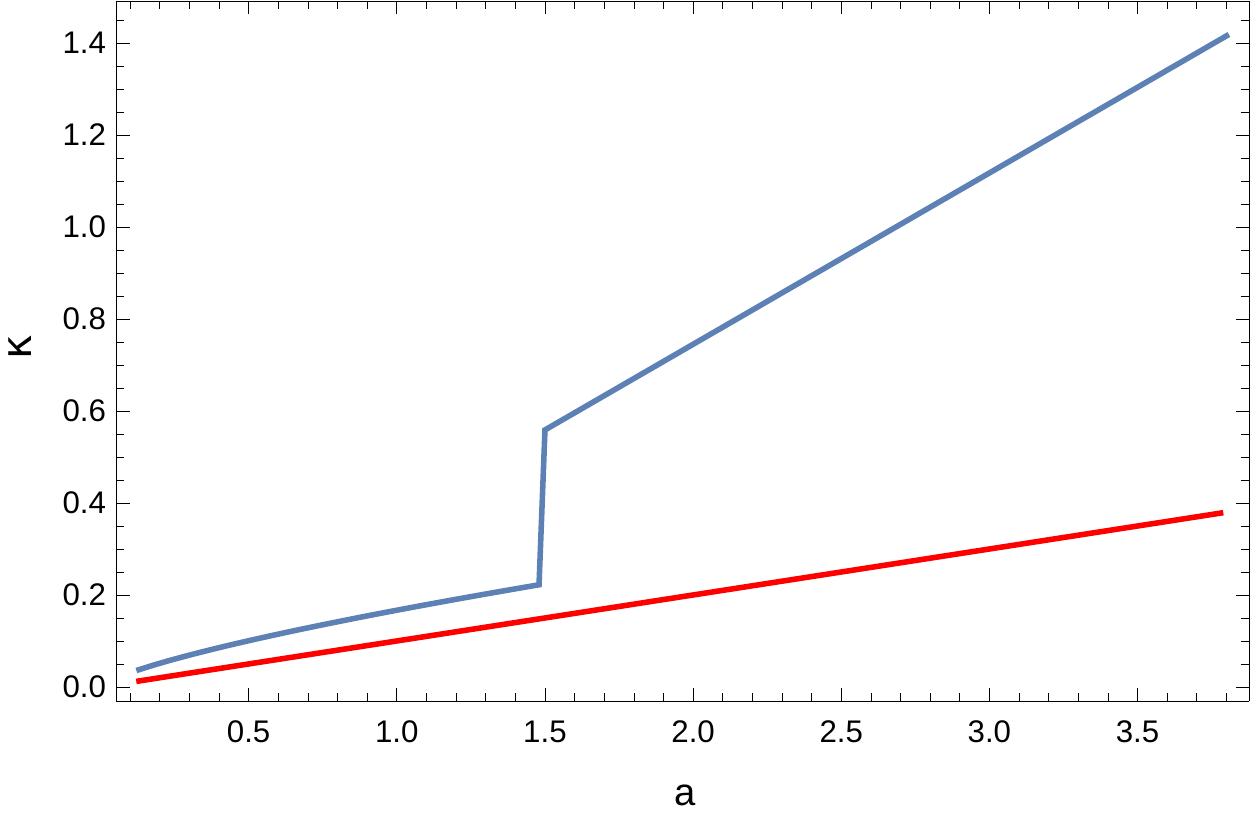}
\caption{{\it Upper left panel:} Free Energy $F$ vs. length dimension $a$ for fixed $T$. {\it Upper right panel:} Pressure $P$ vs. length dimension $a$ for fixed $T$ showing the jump. {\it Lower panel:} Compressibility $\kappa$ vs.  length dimension $a$ for fixed $T$ showing jump. In all the plots the red and blue curves respectively represent the ideal gas and CTC profiles.}
\label{fig:9-11}
\end{center}
\end{figure}

In the next set of graphs, summarized in Fig. \ref{fig:9-11}, we plot the Free Energy as a function of $a$ for fixed $T$, pressure $P=(\partial F)/(\partial a)$  and compressibility (or the inverse of bulk modulus) $\kappa =\{(a\partial^2F)/(\partial a^2)\}^{-1}$ respectively. Once again we find that $F(a)$ is continuous across the critical $a$-value, see the upper left panel of Fig. \ref{fig:9-11}. There is a discontinuity in the pressure $P$ against $a$ graph indicating the possibility of a liquid-gas like transition where below critical $a$ pressure rises much sharply with lowering volume in comparison to the zone above critical $a$ (see the upper right panel of Fig. \ref{fig:9-11}). Consistent with this feature a jump is observed in the profile of compressibility $\kappa $ versus $a$ (see the lower panel of Fig. \ref{fig:9-11}).

\begin{figure}
 	\begin{center}
 		\includegraphics[width=0.4\textwidth]{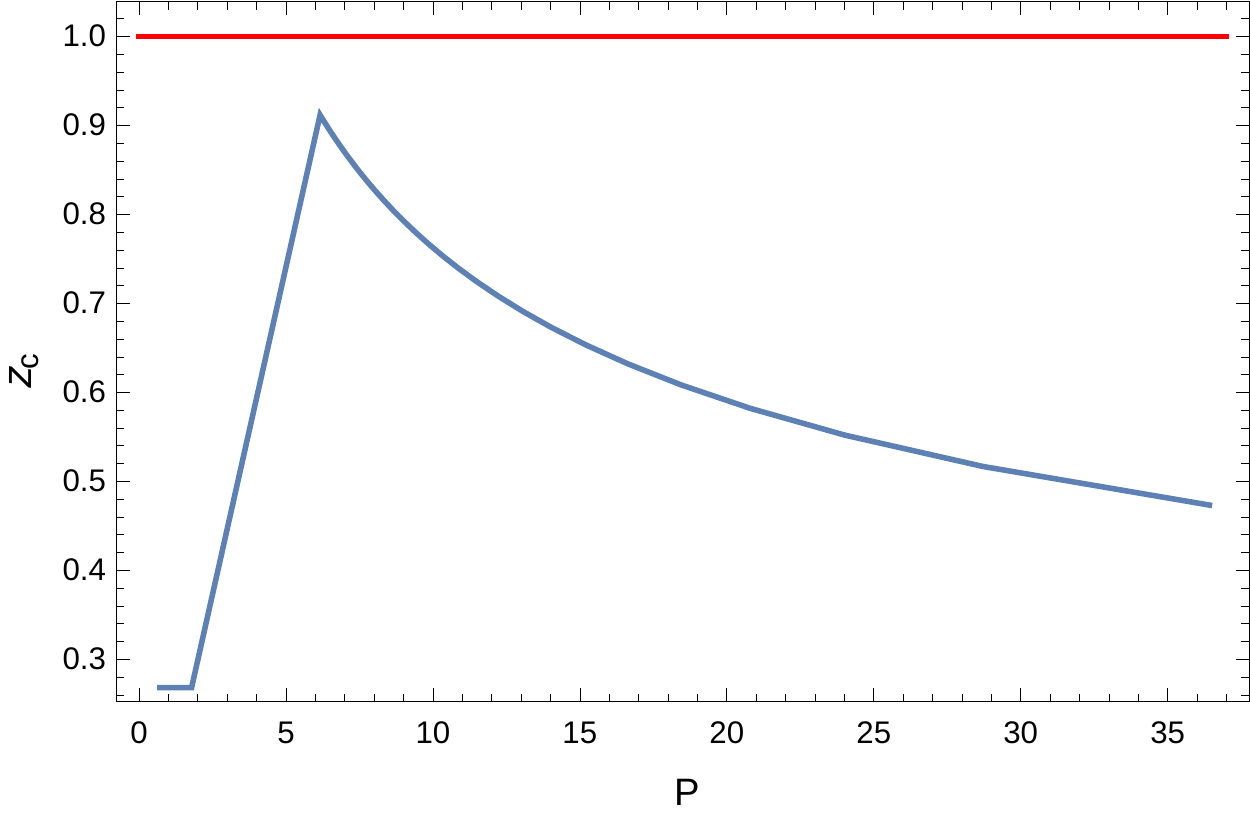}
 		\caption{Plot of the compressibility factor $z_c$ vs. pressure $P$ for the CTC profile (blue curve) showing a deviation from the ideal gas nature (represented by the red curve). }
 		\label{fig12}
 	\end{center}
 \end{figure}

 Lastly in Figure \ref{fig12} we show the behavior of another useful parameter, the  compressibility factor $z_c=(Pa)/T$ against pressure $P$ where $z_{c,ideal}=1$ for ideal gas and any deviation of $z_c$ from unity indicates the non-ideal behavior. Note that $z_{c, CTC}$ peaks to a value close to $1$ for a specific pressure and falls for larger $P$. In conventional non-ideal gases $z$ starts from $1$ for very low pressure (similar to ideal gas) but falls below $1$ as pressure increases and for still higher pressure it goes above the ideal gas value $1$. In CTC $P$ is bounded due to the restriction in $a$.

{\bf {Thermodynamics of multiverse?}} Let us return to cosmology and try to interpret our model of a collection of mini-superspaces as a classical multiverse. Generally multiverse is considered in a quantum framework (see \cite{hart} for a recent perspective, also see \cite{carr}) and its thermodynamic aspects are studied from entanglement entropic point of view \cite{qmulti} (see also \cite{nov}).  In the multiverse picture each of our ``gas'' particles represents a post inflation bubble universe. As suggested by  Tegmark \cite{teg},  the multiverse 
contains  completely  disconnected universes, 
governed by different physical laws or 
mathematical structures. According to Linde \cite{lind} 
 our world (or multiverse) might be a collection of  infinitely
many exponentially large parts and in each of them  different  low-energy laws of physics might prevail. Being very large, each such part evolves effectively independent of other parts. In our model we have considered a simplification where the laws of physics are identical in all member universes. The inflationary scenario is believed to 
provide such a multiverse structure where it is argued that 
our observable domain is just a tiny part
of a single bubble universe that underwent an extra-fast
accelerating phase at some early time
due to the effect of a scalar field. According to Vilenkin \cite{vilenkin1983} and Linde \cite{linde1986}, the concept of ``eternal'' inflation may also lead to the multiverse structure where the universe is continually self-reproducing,
so that an infinite number of bubbles
are extending in both space and time. Proposals for observational signatures 
for multiverse related ideas are in progress \cite{Robles-Perez:2017zrd}.

String landscape\cite{suss} is one of the more popular and concrete version of Multiverse. It has been strongly argued by Mersini-Houghton \cite{mers} that the paradoxical choice of a low entropy ($\sim$ high energy Big Bang inflation) initial condition needs to be addressed as a non-equilibrium dynamic problem. In this framework \cite{mers,mers1}, the ensemble of universes constitute the system that is immersed in a "bath" comprising of very long massive perturbations having super-horizon sized wavelengths (the latter being larger than the horizon size). Effect of   back reaction originating from this bath helps to restrict the choice of initial condition to the apparently unfavorable  (low entropy) one. In this context the present model might be considered as a crude (and brute force scaled down) version of the above with an identical set of universes where the "temperature" might be connected with the collective effect of the environment or bath made up of the super-horizon perturbations. 

In the light of above ideas the present work can be tentatively thought to be a study of classical statistical mechanics of multiverse consisting of a collection of identically behaving mini-superspaces. The roles of different thermodynamic observables along with their discontinuous behavior, leading to the possibility of a phase transition, need to be explored.  

\vspace{0.4cm} 

{\bf{Acknowledgments:}} The authors thank Krishnendu Sengupta for several discussions. We thank Fabrizio Ernesto Canfora, Daniele Musso and Evgeny Novilov for informing us of their works. Also we are grateful to the referee for constructive comments. The work of PD has been supported by INSPIRE, Department of Science and Technology, Govt. of India. SP acknowledges the support through the Faculty Research and Professional Development Fund (FRPDF) Scheme of Presidency University, Kolkata, India.

\appendix

\section{}	

After introducing the Lagrangian multiplier $\lambda$ , the Lagrangian (2) can be written as \cite{tolly},
\begin{eqnarray}\label{lagr}
L=\left( -a \dot{a}^2  + k a - \frac{\Lambda}{3} a^3 \right) + p \left(\frac{\dot{a}^4}{a}+\frac{k^2}{a} \right)+2pk \frac{\dot{a}^2}{a}+ q \dot{a}^2 \ddot{a}+ p a \ddot{a}^2+r \dot{a}^2 \lambda+s a \lambda^2 \pm \sqrt{4 p s}\lambda a \ddot{a}.  
\end{eqnarray}	
where $r$ and $s$ are constants.	
Following \cite{tolly} we vary $\lambda$ in  \ref{lagr} and eliminate it to  rewrite the Lagrangian as,
\begin{eqnarray}
L=\frac{\dot{a}^4}{a}(p-\frac{r^2}{4 s})+\dot{a}^2 \ddot{a}(q\mp r\sqrt{p/s})-a \dot{a}^2+ k a+ p \frac{k^2}{a}+2pk \frac{\dot{a}^2}{a}-\frac{\Lambda}{3} a^3.
\label{lag1}
\end{eqnarray}
Not that total (time-)derivative terms are dropped and furthermore $\dot{a}^2 \ddot{a}$-term also will be dropped being a total derivative term.
The conjugate momentum corresponding to the Lagrangian are,
\begin{eqnarray}
p_1= \frac{\partial L}{\partial \dot{a}}-\frac{d}{dt}(\frac{\partial L}{\partial \ddot{a}})=4 \frac{\dot{a}^3}{a}(p-\frac{r^2}{4s})+4pk\frac{\dot{a}}{a} - 2 a \dot{a} ,
\label{p1}
\end{eqnarray}
\begin{eqnarray}
p_2=\frac{\partial L}{\partial \ddot{a}}=(q\mp r\sqrt{p/s}).
\label{p2}
\end{eqnarray}
Thus the Hamiltonian corresponds to,
\begin{eqnarray}
H&=&p_1 \dot{a}+p_2\ddot{a} -L \nonumber \\
&=&3 \frac{\dot{a}^4}{a}(p-\frac{r^2}{4s})+2pk \frac{\dot{a}^2}{a}-a \dot{a}^2- ka-p \frac{k^2}{a}+\frac{\Lambda}{3} a^3 \nonumber \\
&=& \frac{ 3(p-\frac{\sigma}{4})}{a}\left[\dot{a}^2-\frac{(a^2-2pk)}{6(p-\frac{\sigma}{4})}\right]^2 +V_{eff}
\label{hamil}
\end{eqnarray}
where the effective potential, $V_{eff}=\frac{\Lambda}{3}a^3-p \frac{k^2}{a}-ka-\frac{(a^2-2pk)^2}{12a(p-\frac{\sigma}{4})}$ and $\sigma=\frac{r^2}{s}$.\\

\section{•}
After removing the ghost problems, the compact form of Lagrangian (3) reads
\begin{eqnarray}
L&=&\left( -a \dot{a}^2  + k a - \frac{\Lambda}{3} a^3 \right) + p \left(\frac{\dot{a}^4}{a}+\frac{k^2}{a} \right)-\sigma \frac{\dot{a}^4}{4a}+2pk \frac{\dot{a}^2}{a} \nonumber \\
&=& f(a)\dot{a}^4+g(a) \dot{a}^2+ h(a)
\label{lag2}
\end{eqnarray}
where $f=(p-\frac{\sigma}{4})\frac{1}{a}$ , $g=\frac{2pk}{a}-a$ and $h=p \frac{k^2}{a}+ka-\frac{\Lambda}{3} a^3$.\\
Following \cite{zhao} we write the Lagrangian in terms of a new variable $\rho$ (where we have just use the constraint $\rho=\dot{a}$),
\begin{eqnarray}
L=f\rho^4+g \rho^2+h+\gamma (\rho-\dot{a})
\label{lag3}
\end{eqnarray}
In Hamiltonian framework \cite{dirac} the constraints are,
\begin{eqnarray}
\chi_1=\pi_\rho,~ \chi_2=\pi_\gamma,~ \chi_3=\pi_a+\gamma,~ \chi_4=4f\rho^3+2 g \rho+\gamma 
\label{c1}
\end{eqnarray} 
The constraint matrix $\{\chi_i,\chi_j\}$ ( where $i$ and $j$ goes from 1 to 4) can be written as,
\begin{eqnarray}
\begin{pmatrix}
0 & 0 & 0 & -A  \\
0 & 0 & -1 & -1  \\
0 & 1 & 0 & -B \\
A & 1 & B & 0  
\end{pmatrix}
\end{eqnarray}
where, $A=12 f \rho^2+2g$ and $B=4 f^\prime \rho^3+2 g^\prime \rho$.\\
And the invertible constraint matrix $\{\chi_i,\chi_j\}^{-1}$ reads ($i$ and $j$ goes from 1 to 4),
\begin{eqnarray}
\begin{pmatrix}
0 & B/A  & -1/A & 1/A  \\
-B/A & 0 & 1 & 0  \\
1/A & -1 & 0 & 0 \\
-1/A & 0 & 0 & 0  
\end{pmatrix}
\end{eqnarray}

Thus the Hamiltonian yields,
\begin{eqnarray}
H&=&3f\rho^4+g\rho^2-h \nonumber \\
&=&3f(\rho^2+\frac{g}{6f})^2+V_{eff}
\label{h1}
\end{eqnarray}
where, $V_{eff}=(-h-\frac{g^2}{12f})$. \\ 

Now in a generic Second Class Constraint system with $n$  Second Class Constraints $\chi_i$, $i=1,2,..n$, the modified symplectic structure (or Dirac brackets) is defined in the following way,
\begin{equation}
\{C,D\}^*=\{C,D\}-\{C,\chi _i\}\{\chi ^i,\chi ^j\}^{-1}\{\chi _j,D\}, \label{a3}
\end{equation}
where $\{\chi ^i,\chi ^j\}$ is the  constraint matrix. 

Hence the bracket structure among the variables $\rho$ and $a$ stands
\begin{eqnarray}
\{\rho,a\}=-\frac{1}{A}
\end{eqnarray}
where we have used the definition of Dirac Brackets \cite{dirac}. 
Thus the required symplectic form $d\Omega$ is given by the inverse of the bracket structure,
\begin{eqnarray}
d\Omega=A d\rho da
\end{eqnarray}
and  the partition function reads as,
\begin{eqnarray}
Z&=& \int e^{-\beta H} d\Omega \nonumber \\
&=& e^{h_0} \int (12 f_0\rho^2+2 g_0) e^{-( 3 \beta f_0\rho^4+\beta g_0 \rho^2)} d\rho da
\label{z1}
\end{eqnarray} 
where $\beta=1/T$ and $T$ is the temperature.


\end{document}